\documentclass[nofootinbib,prd,aps]{revtex4}
\usepackage{graphicx}% Include figure files
\usepackage{dcolumn}% Align table columns on decimal point
\usepackage{bm}% bold math

\def \be{\begin{equation}}
\def \ee{\end{equation}}

\begin{document}
\voffset = 0.3 true in

\title{A Canonical $D_s(2317)$?}

\author{Olga Lakhina}
\affiliation{
Department of Physics and Astronomy, University of Pittsburgh,
Pittsburgh PA 15260}

\author{Eric S. Swanson}
\affiliation{
Department of Physics and Astronomy, University of Pittsburgh,
Pittsburgh PA 15260}

\begin{abstract}
\vskip .3 truecm
It is shown that quark mass dependence induced by one loop corrections to the Breit-Fermi
spin-dependent one gluon exchange potential permits an accurate determination of heavy-light
meson masses. Thus the $D_s(2317)$ is a canonical $c\bar s$ meson in this scenario. 
The multiplet splitting relationship of chiral doublet models, $M(1^+)-M(1^-) = M(0^+) -M(0^-)$,
holds to good accuracy in the $D$ and  $D_s$ systems, but is accidental. Radiative transitions 
and 
bottom flavoured meson masses are discussed.
\end{abstract}

%\pacs{}
\maketitle

\section{Introduction}  % ----------------------------------------

BaBar's discovery of the $D_s(2317)$ state\cite{BabarDs} generated strong interest in heavy meson
spectroscopy, chiefly due to its surprisingly low mass with respect to expectations.
These expectations are based on quark models or lattice gauge theory. Unfortunately,  at present
large lattice
systematic errors do not allow a determination of the $D_s$ mass with  a precision better than
several hundred MeV. And, although quark models appear to be exceptionally accurate in describing
charmonia, they are less constrained by experiment and on a weaker theoretical footing in the
open charm sector. It is therefore imperative to examine reasonable alternative descriptions of
the open charm sector.

The $D_s(2317)$ was produced in $e^+e^-$ scattering and discovered in the isospin
violating $D_s\pi$ decay mode in $K\bar K \pi\pi$ and $K\bar K \pi\pi\pi$ mass distributions. Its
width is less than 10 MeV and it is likely that the quantum numbers are $J^P = 0^+$\cite{HHreview}.
Finally, if the $D_s\pi^0$ mode dominates the width of the $D_s(2317)$ then the measured 
product of branching ratios\cite{belleDs} 
\be
Br(B^0 \to D_{s}(2317)K) \cdot Br(D_{s}(2317) \to D_s \pi^0) = (4.4 \pm 0.8 \pm 1.1)\cdot 10^{-5}
\ee
implies that $Br(B \to D_{s}(2317)K) \approx Br(B \to D_s K)$, consistent with the $D_s(2317)$ being
a canonical $0^+$ $c\bar s$ meson.

In view of this,
Cahn and Jackson have examined the feasibility of describing the masses and decay widths of the
low lying $D$ and $D_s$ states within the constituent quark model\cite{CJ}. They assume a 
standard spin-dependent structure for the quark-antiquark interaction (see below) and allow
general vector and scalar potentials. Their conclusion is that it is very difficult to describe
the data in this scenario.

Indeed, the $D_s(2317)$ lies some 160 MeV below most model predictions (see Ref.\cite{HHreview} 
for a summary), leading to speculation that the state could be a $DK$ molecule\cite{BCL} or a 
tetraquark\cite{tetra}. Such speculation  is supported by the isospin violating discovery mode
of the $D_s(2317)$ and the proximity of the S-wave $DK$ threshold at 2358-2367 MeV.  Other
studies have been made with QCD sum rules\cite{narison}, using heavy quark symmetry to examine 
decay models\cite{cdf}, or in unitarised chiral models\cite{guo}.

Although these proposals have several attractive features, it is important to exhaust 
possible canonical $c\bar s$ descriptions of the $D_s(2317)$
before resorting to more exotic models. 
%In view of this we examine the implications of a spin-dependent
%interaction that is motivated by 
Here we propose a simple modification to the standard vector Coulomb+scalar linear quark potential
model that maintains good agreement with the charmonium spectrum and agrees remarkably
well with the $D$ and $D_s$ spectra. Possible experimental tests of this scenario are discussed.

\section{A Quark Model of Open Charm States}

The spectra we seek to explain are summarised in Table \ref{DDsSpectraTab}.  Unfortunately, the
masses of 
the $D_0$ (labelled $a$) and $D_1$ (labelled $b$) are poorly determined. Belle have observed\cite{belleD0} the 
$D_0$ in $B$ decays and
claim a mass of $2308  \pm 17 \pm 32$ MeV with a width of $\Gamma = 276 \pm 21 \pm 18 \pm 60$, while FOCUS\cite{focusD0} find $2407 \pm 21 \pm 35$ MeV with a width $\Gamma = 240 \pm 55 \pm 59$. 
While some
authors choose to average these values, we regard them as incompatible and consider the cases
separately below. Finally, there is an older mass determination from Belle\cite{belleOld} of 
$2290 \pm 22 \pm 20$ MeV with a width of
$\Gamma = 305 \pm 30 \pm 25$.  
The $D_1'$ has been seen in $B$ decays to $D\pi\pi$ and $D^*\pi\pi$  by Belle\cite{belleD1}. A Breit-Wigner fit yields a mass of $2427 \pm 26 \pm 20 \pm 15$ MeV and a width of $384^{+107}_{-90}\pm 24 \pm 70$ MeV. Alternatively, a preliminary report from 
CLEO\cite{cleoD1} cites a mass of $2461^{+41}_{-34} \pm 10 \pm 32$ MeV and a width of $290^{+101}_{-79}\pm 26 \pm 36$ MeV. Finally, FOCUS\cite{focusD1} obtain a lower neutral $D_1'$ mass of $2407 \pm 21 \pm 35$ MeV. Other masses in Table \ref{DDsSpectraTab} are obtained from the PDG compilation\cite{PDG}.

%$\Gamma(D_2) = 46.4 \pm 4.4 \pm 3.1$

\begin{table}[h]
\caption{Low Lying $D$ and $D_s$ Spectra (MeV)}
\label{DDsSpectraTab}
\begin{tabular}{l|cccccc}
\hline
\hline
$J^P$ & $0^-$ &  $1^-$ & $0^+$  &  $1^+$  &  $1^+$  & $2^+$ \\
\hline
$D$ & $1869.3 \pm 0.5$ & $2010.0 \pm 0.5$ & a  & b & $2422.2 \pm 1.8$  & $2459 \pm 4$ \\
$D_s$ & $1968.5 \pm 0.6$ & $2112.4 \pm 0.7$ &  $2317.4 \pm 0.9$ & $2459.3 \pm 1.3$ &  $2535.35 \pm 0.34$ & $2572.4 \pm 1.5$ \\
\hline
\end{tabular}
\end{table}

In addition to the unexpectedly low mass of the $D_s(2317)$, the $D_s(2460)$ is also somewhat 
below predictions assuming it is the $D_{s1}$ (Godfrey and Isgur, for example,  predict a mass of 2530 MeV\cite{GI}).
It is possible that an analogous situation holds in the $D$ spectrum, 
depending on the mass of the $D_0$.
The quark model explanation of these states rests on P-wave mass splittings induced by spin-dependent
interactions. A common model of spin-dependence is based on the Breit-Fermi reduction of the 
one-gluon-exchange interaction supplemented with the spin-dependence due to a scalar current 
confinement interaction. The general form of this potential has been 
computed by Eichten and Feinberg\cite{EF} at tree level using Wilson loop methodology. The
result is parameterised in terms of four nonperturbative matrix elements, $V_i$, which can be determined
by electric and magnetic
field insertions on quark lines in the Wilson loop.
Subsequently, Gupta and Radford\cite{PTN} performed a one-loop computation of the heavy quark 
interaction and showed that a fifth interaction, $V_5$ is present in the case of unequal quark
masses. The net result is a quark-antiquark interaction that can be written as:

\begin{eqnarray}
V_{q\bar q}=V_{conf}+V_{SD}
\end{eqnarray}
where $V_{conf}$ is the standard Coulomb+linear scalar form:
\begin{equation}
V_{conf}(r)=-\frac{4}{3}\frac{\alpha_s}{r}+br
\end{equation} 
and

\begin{eqnarray}
V_{SD}(r) &=& \left( {\bm{\sigma}_q \over 4 m_q^2} +
{\bm{\sigma}_{\bar q} \over 4 m_{\bar q}^2} \right)\cdot {\bf L} \left( {1\over r}
{d V_{conf} \over d r} + {2 \over r} {d V_1 \over d r} \right) +
\left( {\bm{\sigma}_{\bar q} +
        \bm{\sigma}_q \over 2 m_q m_{\bar q}} \right)\cdot {\bf L}
        \left( {1 \over r} {d V_2 \over d r} \right) \nonumber \\
&&+ {1 \over 12 m_q m_{\bar q}}\Big( 3 \bm{\sigma}_q \cdot \hat {\bf r} \,
 \bm{\sigma}_{\bar q}\cdot \hat {\bf r} -   \bm{\sigma}_q\cdot
 \bm{\sigma}_{\bar q} \Big) V_3
+ {1 \over 12 m_q m_{\bar q}} \bm{\sigma}_q \cdot \bm{\sigma}_{\bar q}
V_4 \nonumber \\
&&+ {1\over 2}\left[ \left( {\bm{\sigma}_q \over m_q^2} - {\bm{\sigma}_{\bar q}\over m_{\bar q}^2}\right)\cdot {\bf L} +
\left({\bm{\sigma}_q - \bm{\sigma}_{\bar q}\over m_q m_{\bar q}}\right)\cdot {\bf L} \right] V_5.
\label{VSD}
\end{eqnarray}
Here ${\bf L} = {\bf L}_q = - {\bf L_{\bar q}}$,
$r=|{\bf r}|= |{\bf r}_q - {\bf r}_{\bar q}|$ is the ${\bar Q Q}$ separation
and the $V_i=V_i(m_q,m_{\bar q}; r)$ are the Wilson loop matrix elements discussed above.

The first four $V_i$ are order $\alpha_s$ in perturbation theory, while $V_5$ is order 
$\alpha_s^2$; for this
reason $V_5$ has been largely ignored by quark modellers. 
The exceptions known to us are Ref. \cite{GRR}, which  examines S-wave masses for a variety of heavy-light mesons in a model very similar to that presented here, and the second of Ref. \cite{PTN}, which
does not consider scalar confinement contributions to the spin-dependent interaction.
More recently, Cahn and Jackson\cite{CJ}
only consider $V_1$ -- $V_4$ in an analysis of the $D_s$ system.
In practice this is acceptable (as we show below) {\it except in the case
of unequal quark masses}, where the additional spin-orbit interaction can play an important role.

Here we propose to take the spin-dependence of Eqn. \ref{VSD} seriously and examine its effect
on low-lying heavy-light mesons. Our model can be described in terms of vector and scalar kernels 
defined by

\be
V_{conf} = V + S
\ee
where
$V = -4\alpha_s/ 3 r$ is the vector kernel and $S = br$ is the scalar kernel, and by the 
order  $\alpha_s^2$ contributions to the $V_i$, denoted by $\delta V_i$.
Expressions for the matrix elements of the spin-dependent interaction are then 

\begin{eqnarray}
V_1 &=& -S + \delta V_1 \nonumber \\
V_2 &=& V + \delta V_2 \nonumber \\
V_3 &=& V'/r - V'' + \delta V_3 \nonumber \\
V_4 &=& 2 \nabla^2 V + \delta V_4\nonumber \\
V_5 &=& \delta V_5.
\end{eqnarray}
Explicitly, 

\begin{eqnarray}
V_1(m_q,m_{\bar{q}},r)&=&-br-C_F\frac{1}{2r}\frac{\alpha_s^2}{\pi}
\left(C_F-C_A \left(\ln{\left[(m_qm_{\bar{q}})^{1/2}r\right]}+\gamma_E\right)\right)\nonumber\\
V_2(m_q,m_{\bar{q}},r)&=&-\frac{1}{r}C_F\alpha_s\left[1+\frac{\alpha_s}{\pi}
\left[\frac{b_0}{2}[\ln{(\mu r)}+\gamma_E]+\frac{5}{12}b_0-\frac{2}{3}C_A+
\frac{1}{2}\left(C_F-C_A \left(\ln{\left[(m_qm_{\bar{q}})^{1/2}r\right]}+\gamma_E\right)\right)\right]\right]\nonumber\\
V_3(m_q,m_{\bar{q}},r)&=&\frac{3}{r^3}C_F\alpha_s\left[1+\frac{\alpha_s}{\pi}
\left[\frac{b_0}{2}[\ln{(\mu r)}+\gamma_E-\frac{4}{3}]+\frac{5}{12}b_0-\frac{2}{3}C_A+\right.\right.\nonumber\\
&&+\left.\left.\frac{1}{2}\left(C_A+2C_F-2C_A \left(\ln{\left[(m_qm_{\bar{q}})^{1/2}r\right]}+\gamma_E-\frac{4}{3}\right)\right)\right]\right]\nonumber\\
V_4(m_q,m_{\bar{q}},r)&=&\frac{32\alpha_s\sigma^3 e^{-\sigma^2 r^2}}{3\sqrt{\pi}}\nonumber\\
V_5(m_q,m_{\bar{q}},r)&=&\frac{1}{4r^3}C_FC_A\frac{\alpha_s^2}{\pi}\ln{\frac{m_{\bar{q}}}{m_q}}
\label{Vmodel}
\end{eqnarray}
where $C_F=4/3$, $C_A=3$, $b_0=9$, $\gamma_E=0.5772$, and the scale $\mu$ has been set to 1 GeV.

The hyperfine interaction (proportional to $V_4$) contains a delta function in configuration space
and is normally `smeared' to make it nonperturbatively tractable. For this reason we choose not to
include $\delta V_4$ in the model definition of Eqn. \ref{Vmodel}.
In the following, the hyperfine interaction has been treated nonperturbatively 
and the remaining spin-dependent terms are evaluated in leading-order 
perturbation theory.

We have confirmed that the additional features do not ruin previous agreement with
the charmonium spectrum. For example, Ref. \cite{bgs} obtains very good agreement with experiment
for parameters $m_c = 1.4794$ GeV, $\alpha_s = 0.5461$, $b = 0.1425$ GeV$^2$, and $\sigma = 1.0946$ GeV.
Employing the model of Eqn. \ref{Vmodel} worsens the agreement with experiment, but the original good
fit is recovered upon slightly modifying parameters (the refit parameters are $m_c = 1.57$ GeV, $\alpha_s = 0.52$, $b = 0.15$ GeV$^2$, and $\sigma = 1.3$ GeV).

\begin{table}[h]
\caption{Model Parameters}
\label{ParamsTab}
\begin{tabular}{l|ccccc}
\hline
\hline
model & $\alpha_s$ & $b$ (GeV$^2$) & $\sigma$ (GeV) & $m_c$ (GeV) & $C$ (GeV) \\
\hline
low & 0.46 & 0.145 & 1.20 & 1.40 & -0.298 \\
avg & 0.50 & 0.140 & 1.17 & 1.43 & -0.275 \\
high & 0.53 & 0.135 & 1.13 & 1.45 & -0.254 \\
\hline
\end{tabular}
\end{table}

The low lying  $c\bar s$ and $c \bar u$ states are fit reasonably well with the parameters labelled
`avg' in Table \ref{ParamsTab}.  Predicted masses are given in Table \ref{DTab}. Parameters labelled
`low' in Table \ref{ParamsTab} fit the $D$ mesons very well, whereas those labelled `high' fit the
known $D_s$ mesons well. It is thus reassuring that these parameter sets are reasonably similar
to each other and to the refit charmonium parameters. (Note that constant shifts in each flavour 
sector are determined by the relevant pseudoscalar masses.)

The predicted $D_{s0}$ mass is 2341 MeV, 140 MeV lower than the prediction of Godfrey and Isgur and
only 24 MeV higher than experiment.
We remark that the best fit to the $D$ spectrum predicts a mass of 2287 MeV for the $D_0$ meson,
in good agreement with the preliminary Belle measurement of 2290 MeV, 21 MeV below the current Belle 
mass, and in disagreement with the FOCUS mass of 2407 MeV.

The average error in the predicted P-wave masses is less than 1\%.
It thus appears likely that a simple modification to the spin-dependent quark interaction is
capable of describing heavy-light mesons with reasonable accuracy.

\begin{table}[h]
\caption{Predicted Low Lying Charm Meson Spectra (GeV)}
\label{DTab}
\begin{tabular}{l|cccccc}
\hline
\hline
flavour & $0^-$ & $1^-$ & $0^+$ & $1^+$ & $1^+$ & $2^+$ \\
\hline
$D$ & 1.869 & 2.017 & 2.260 & 2.406 & 2.445 & 2.493 \\
\hline
$D_s$ & 1.968  & 2.105 & 2.341 & 2.475 & 2.514 & 2.563 \\
\hline
\end{tabular}
\end{table}

\begin{figure}[h]
\includegraphics[width=4 true cm, angle=270]{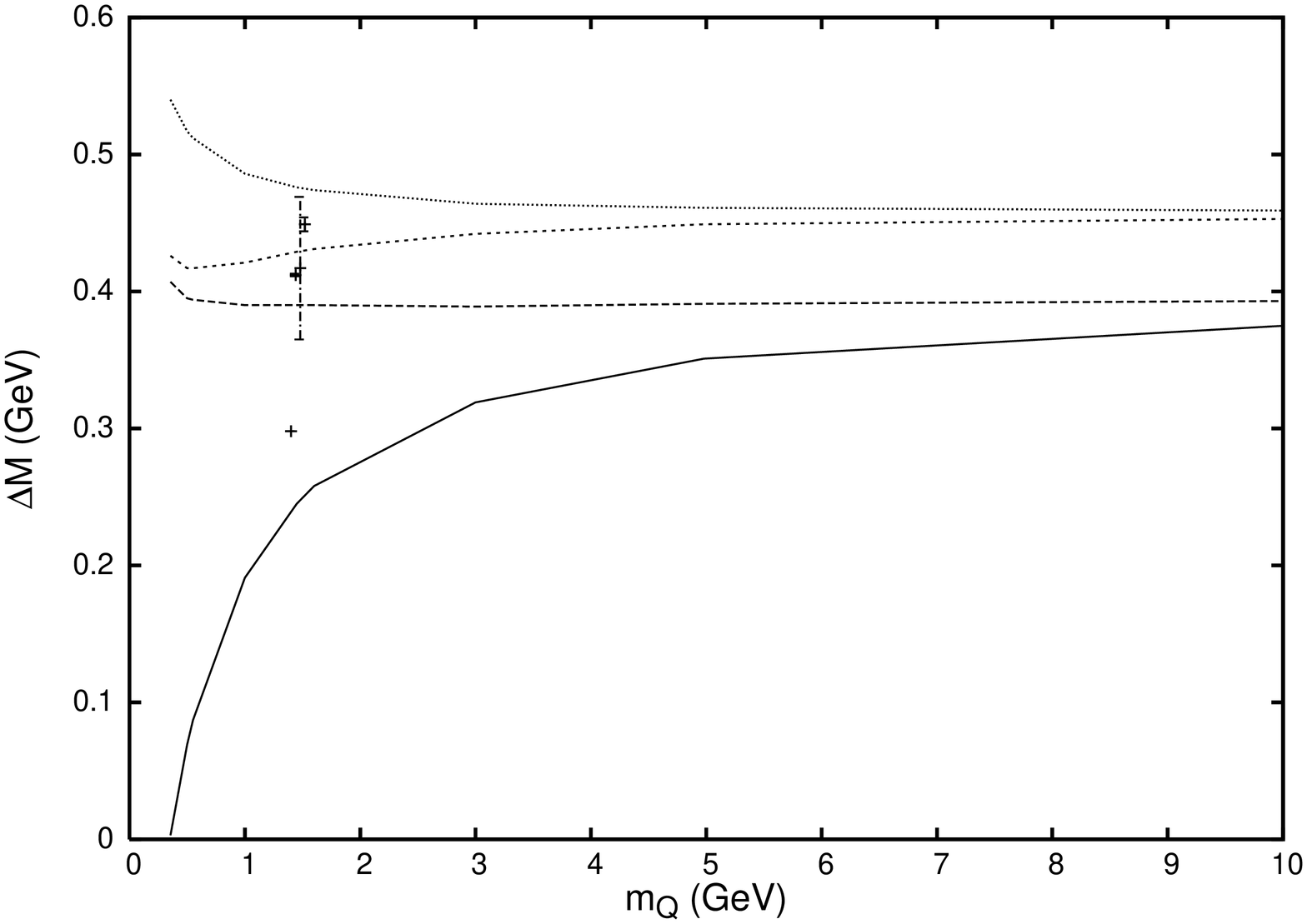}
\hskip 1 true cm
\includegraphics[width=4 true cm, angle=270]{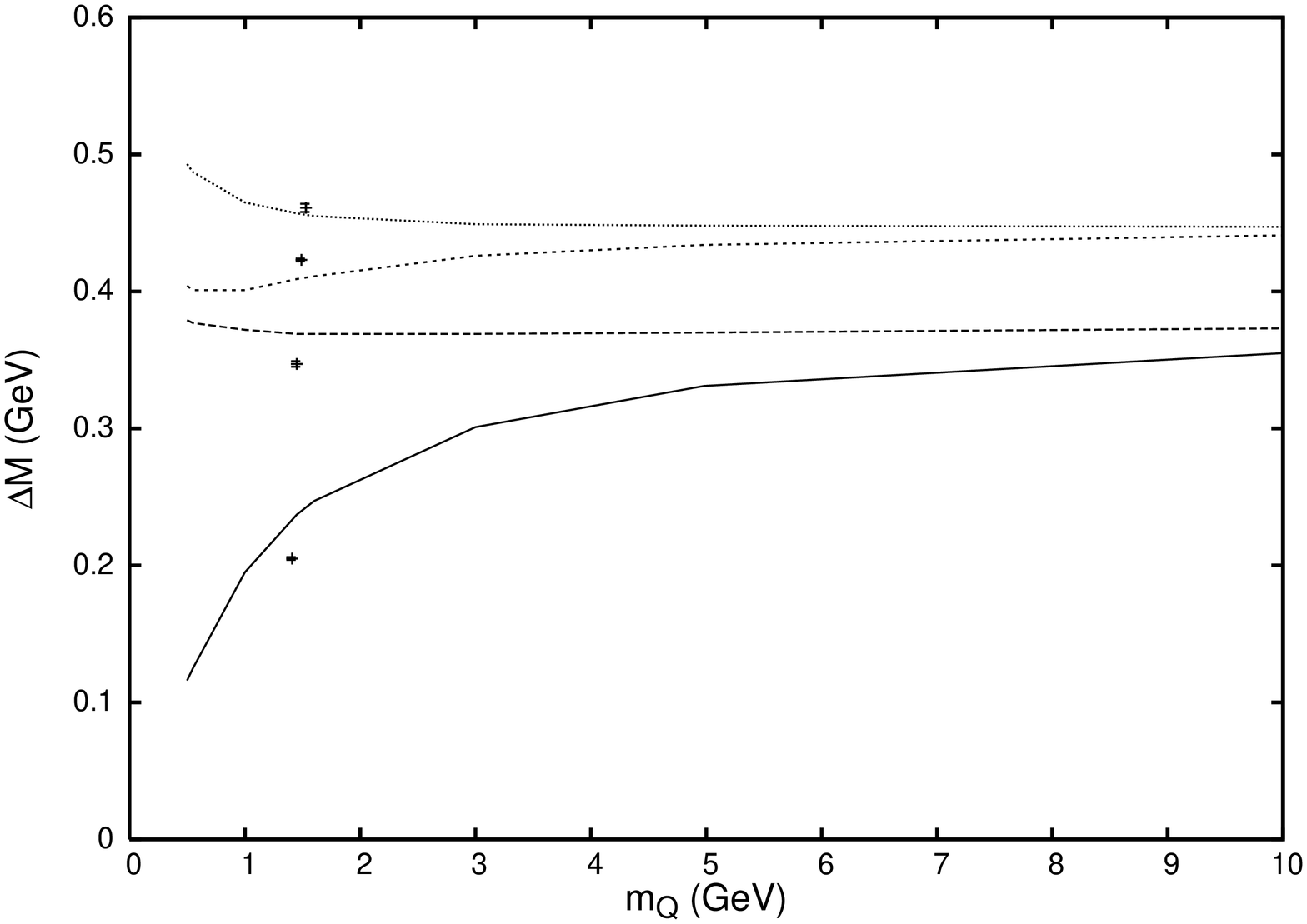}
\includegraphics[width=4 true cm, angle=270]{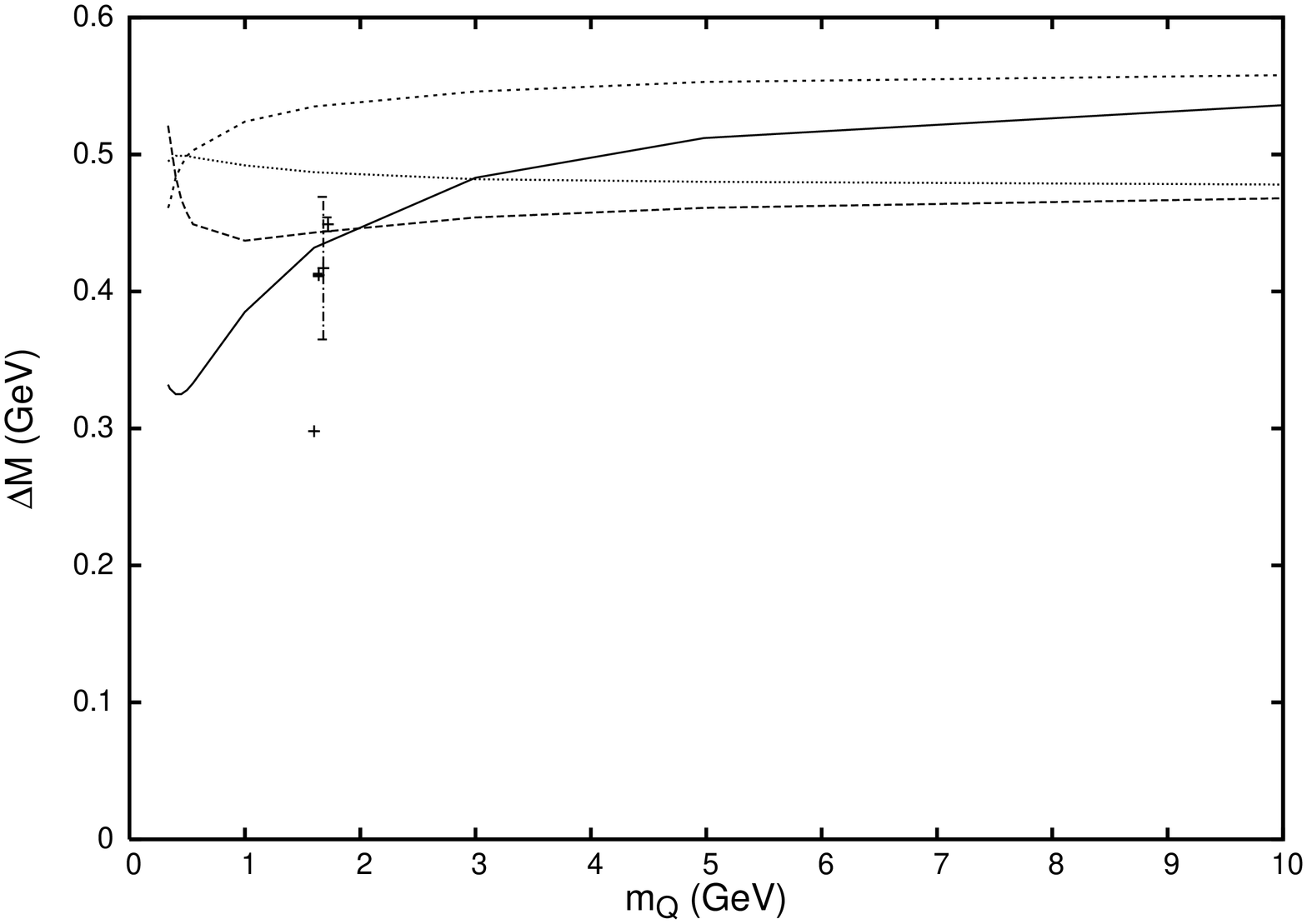}
\hskip 1 true cm
\includegraphics[width=4 true cm, angle=270]{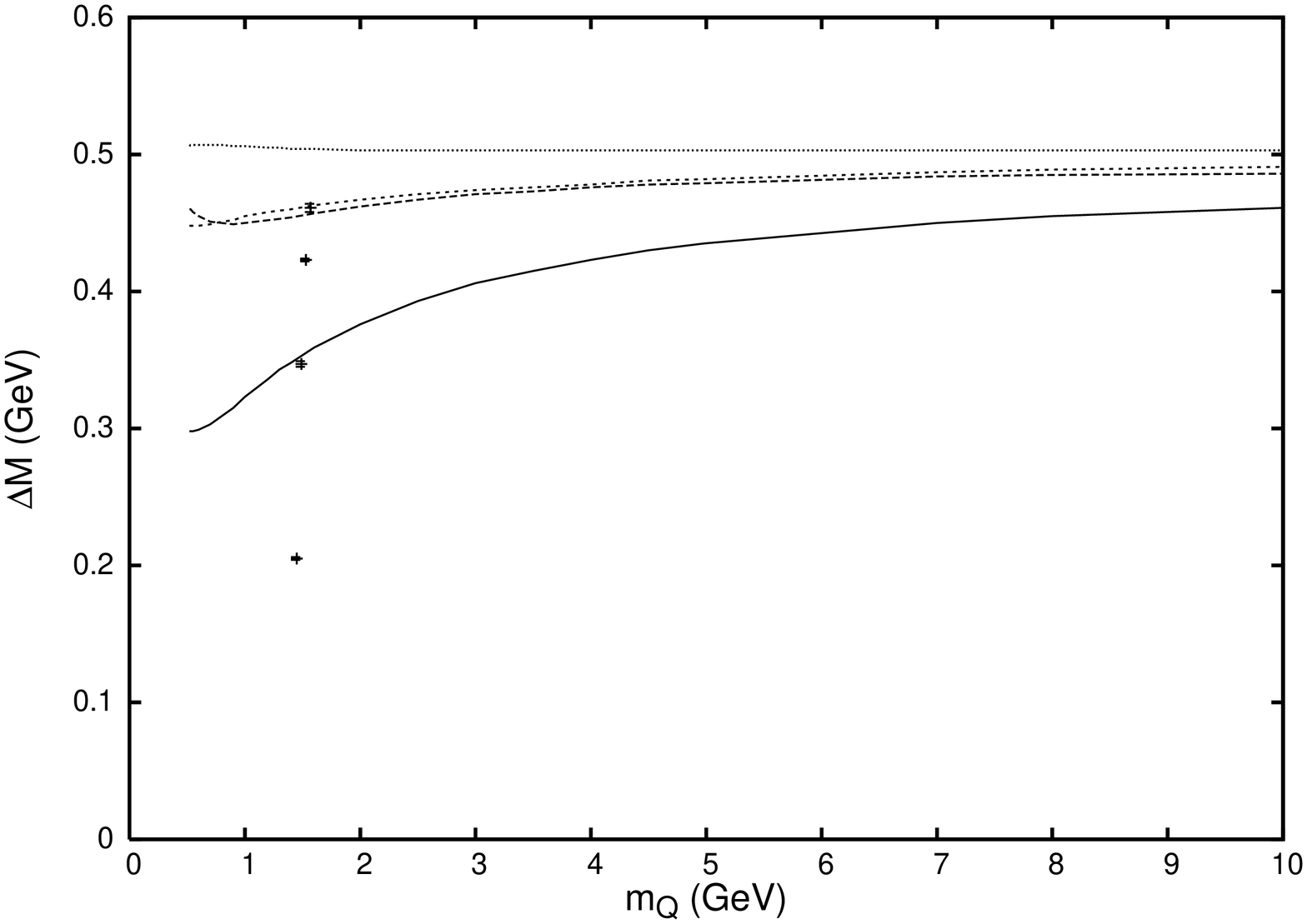}
\caption{M(P-wave) - M(vector) as a Function of the Heavy Quark Mass. $D$ System (left); $D_s$ System (right); one-loop model (top); traditional model (bottom). $D_2$ (dotted line); $D_1'$ (thin dashed line); $D_1$ (dashed line); $D_0$ (solid line). The points are experimental mass splittings.}
\label{massesFig}
\end{figure}

We examine the new model in more detail by computing P-wave meson masses (with respect to the ground
state vector) as a function of the heavy quark mass. Results for $Q\bar u$ and $Q\bar s$ systems
are displayed in Fig. \ref{massesFig}. All panels indicate that the approach to the heavy quark
limit is very slow. In the case of the traditional $Q\bar u$ system the heavy quark doublets
are inverted (with the $j_q=1/2$ doublet higher than the $j_q=3/2$), in disagreement with 
experiment. Alternatively, the one-loop model displays the expected heavy quark behaviour. 
Furthermore, the predicted mass splittings at the charm quark scale are near experiment
for $D$ masses (points on the panels). A similar situation holds for the $D_s$ system
(right panels), except in this case it is the $D_s$ and $D_s'$ that do not agree with the
traditional model predictions. 

Although the reliability of the model is suspect in the case of light $Q$ masses, it is
intriguing that the one-loop model scalar-vector mass difference gets small in this limit.
Thus it is  possible that the enigmatic $a_0$ and $f_0$ mesons may simply
be $q\bar q$ states. 

Finally, one obtains $M(h_c) > M(\chi_{c1})$ in one-loop and traditional models, in agreement
with experiment.  However, experimentally  $M(f_1) - M(h_1) \approx 100$ MeV and $M(a_1)-M(b_1) \approx
0$ MeV, indicating that the $^3P_1$ state is heavier than (or nearly degenerate with) the $^1P_1$ light
meson state. Thus the sign of the combination of tensor and spin-orbit terms that drives this
splitting must change when going from charm quark to light quark masses.  This change is approximately 
correctly reproduced in the traditional model (lower left panel of Fig. \ref{massesFig}). 
The one-loop model does not reproduce the desired cross over, although it does come close, and 
manipulating model parameters can probably reproduce this behaviour. We do not pursue this here
since the focus is on heavy-light mesons.

\section{Mixing Angles and Radiative Decays}

The lack of charge conjugation symmetry implies that two nearby low lying axial vector states 
exist (generically denoted as $D_1$ and $D_1'$ in the following). The mixing angle between these
states can be computed and compared to experiment (with the help of additional model assumptions).
We define the mixing angle via the relations:

\begin{eqnarray}
&|D_1\rangle  &= +\cos(\phi) |^1P_1\rangle + \sin(\phi) | ^3P_1\rangle \nonumber \\
&|D_1'\rangle  &= -\sin(\phi) |^1P_1\rangle + \cos(\phi) | ^3P_1\rangle.
\label{D1Eqn}
\end{eqnarray}

In the following,
we choose to define the $D_1'$ as the heavier axial state in the heavy quark limit.
In this limit
%In the heavy-quark limit 
a particular mixing angle follows
from the quark mass dependence of the spin-orbit and
tensor terms, $\phi_{HQ} = -54.7^o$ $(35.3^o)$,
if the expectation of the heavy-quark spin-orbit interaction
is positive (negative). It is often assumed that the heavy quark mixing angle holds for charmed
mesons.

\begin{figure}[h]
\includegraphics[width=4 true cm, angle=270]{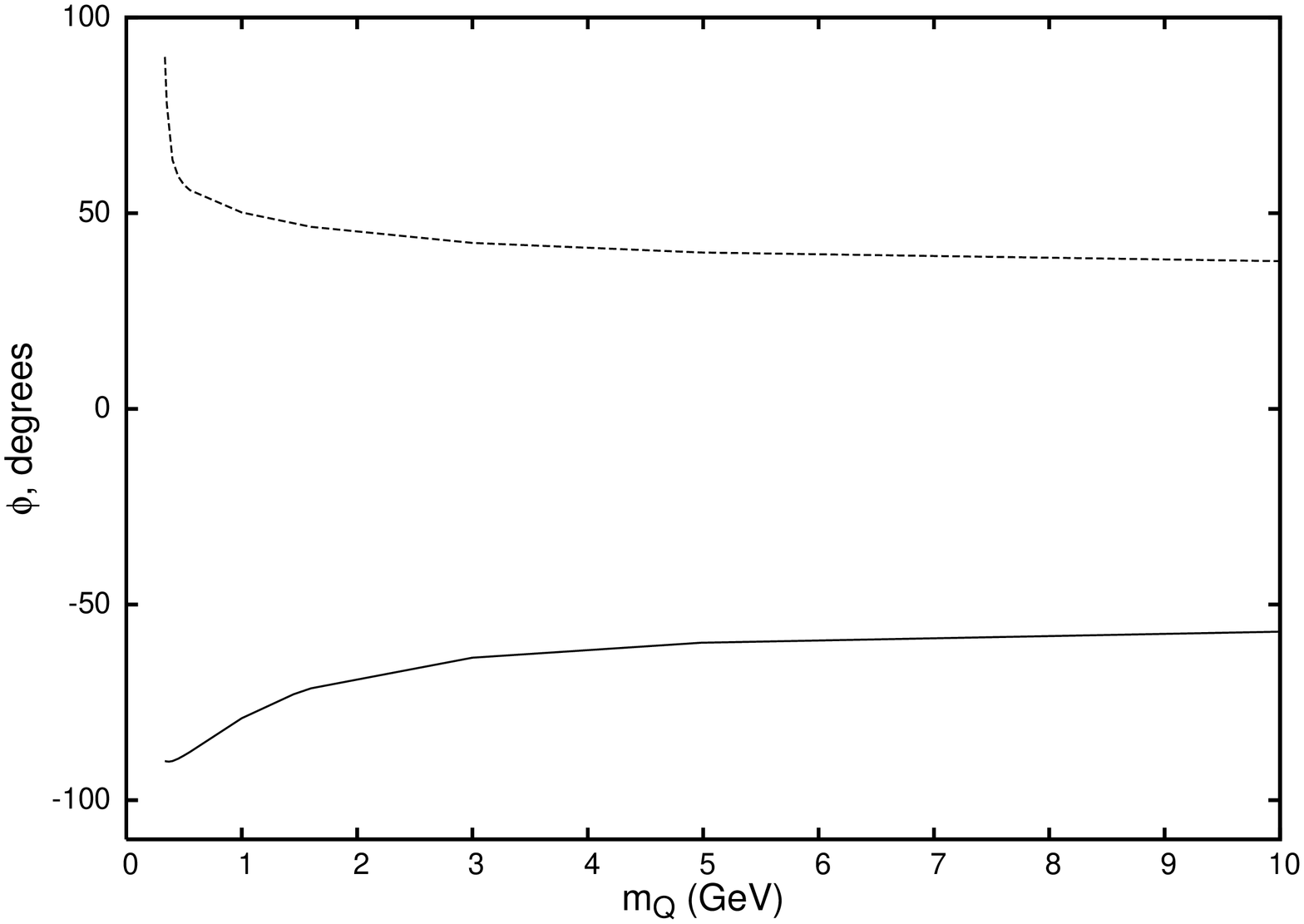}
\hskip 1 true cm
\includegraphics[width=4 true cm, angle=270]{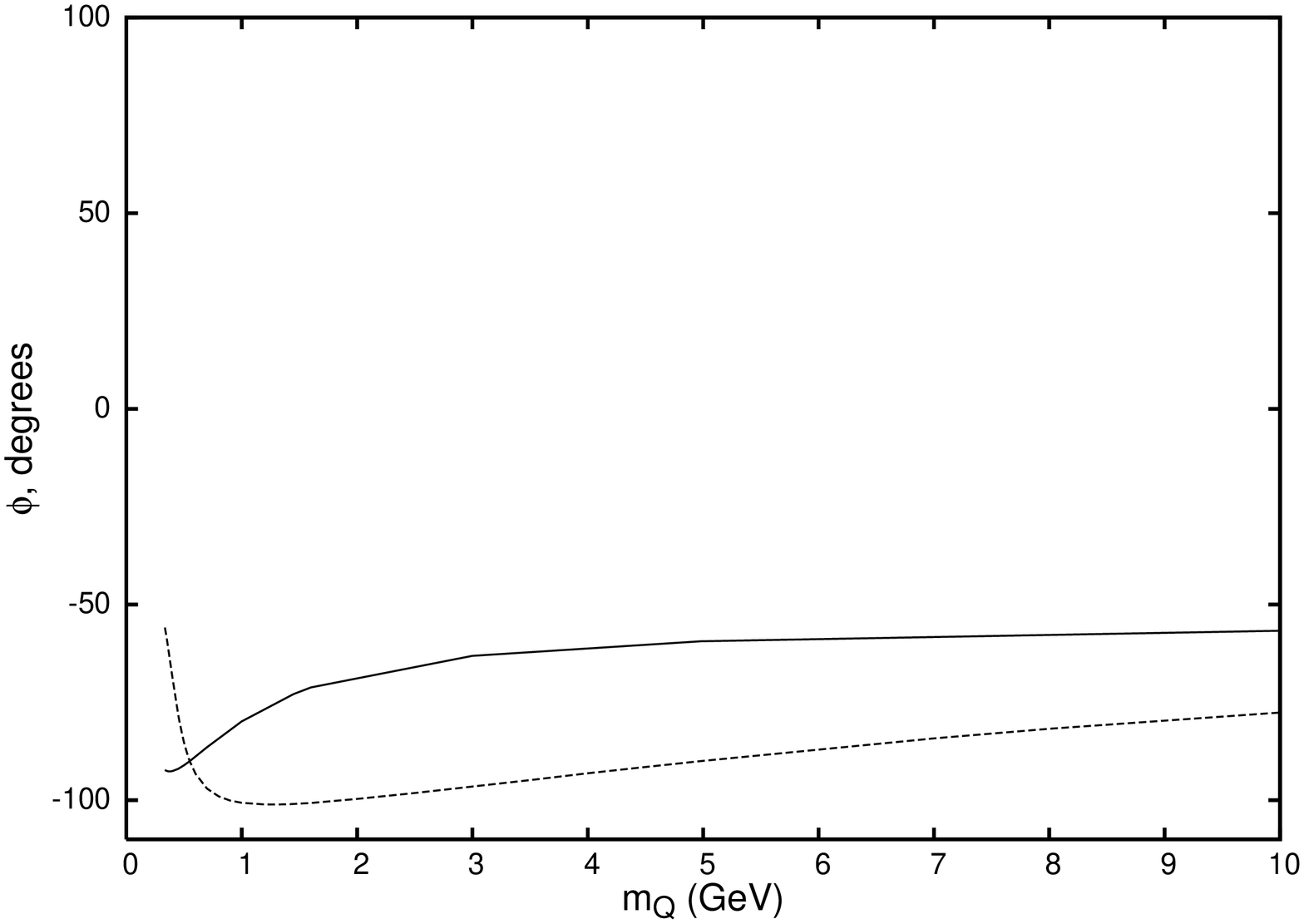}
\caption{$D_1$ (left) and $D_{s1}$ (right) Mixing Angles vs. Heavy Quark Mass. The traditional model is given by the dashed line; the extended model is the solid line.}
\label{mixingAngleFig}
\end{figure}

Fig. \ref{mixingAngleFig} shows the dependence of the mixing angle on the heavy
quark mass for $Q\bar u$ and $Q \bar s$ mesons for the traditional and extended models. 
The effect of the one-loop terms is dramatic: for the $Q\bar u$ system the relevant spin-orbit matrix element changes
sign, causing the heavy quark limit to switch from $35.3^o$ to $-54.7^o$.  Alternatively, both models
approach $-54.7^o$ in the $Q\bar s$ system. 
There is strong deviation from
the heavy quark limit in both cases: $\phi(D_s) \approx \phi(D) \approx - 70^o$. 
This result is not close to the heavy quark limit (which is approached very slowly) -- indeed
it is reasonably close to the unmixed limit of $\pm 90^o$!

Mixing angles can be measured with the aid of strong or radiative decays. For example, the $D_1'$ is
a relatively narrow state, $\Gamma(D_1') = 20.4 \pm 1.7$ MeV, while the $D_1$ is very broad. 
This phenomenon is expected in the heavy quark limit of the $^3P_0$ and Cornell strong decay 
models\cite{HHreview,CS,GK}. Unfortunately, it is difficult to exploit these widths to measure the mixing
angle because strong decay models are rather imprecise.

Radiative decays are possibly more accurate probes of mixing angles because
the decay vertex is established and the impulse approximation has a long history of success.
Table \ref{RadTransTab} presents the results of two computations of radiative decays of $D$ and $D_s$
mesons. Meson wavefunctions are computed with `average' parameters, as above.
Transition matrix elements are evaluated in the impulse approximation and full recoil is allowed.
The column labelled `nonrel' reports transition matrix elements computed in the nonrelativistic limit,
while the column labelled `rel' contains results obtained with the full spinor structure at 
the photon vertex.

The nonrelativistic results can differ substantially from those of Refs. \cite{Godfrey, CS} because
those computations were made in the zero recoil limit where an E1 transition, for example, is
diagonal in spin. Thus the decay $D_1 \to D^* \gamma$ can only proceed via the $^3P_1$ component
of the $D_1$. Alternatively, the computations made here are at nonzero recoil and hence permit
both components of the $D_1$ to contribute to this decay. The table entries indicate that 
nonzero recoil effects can be surprisingly large.

\begin{table}[h]
\caption{Meson Radiative Decay rates (keV).}
\label{RadTransTab}
\begin{center}
\begin{tabular}{|c|c|c|c|c|}
\hline
\hline
mode & $q_\gamma$ (MeV) & nonrel & rel & expt \\
\hline
$D^{+*}\rightarrow\gamma D^+$       &136 &1.38                          &0.08                                &$1.5\pm0.5$\\
$D^{0*}\rightarrow\gamma D^0$       &137 &32.2                          &13.3                                &$<800$ \\
$D_0^+\rightarrow\gamma D^{*+}$     &361 &76.0                          &7.55                                & \\
$D_0^0\rightarrow\gamma D^{*0}$     &326 &1182                          &506.                                & \\
$D_1^+\rightarrow\gamma D^{*+}$     &381 &$(6.34s)^2+(3.22s+5.9c)^2$    &$(2.00s-0.13c)^2+(0.13s+4.23c)^2$   & \\ 
$D_1^0\rightarrow\gamma D^{*0}$     &380 &$(27.05s)^2+(19.33s+9.63c)^2$ &$(17.65s-0.15c)^2+(12.28s+6.01c)^2$ & \\
$D_1'^+\rightarrow\gamma D^{*+}$    &381 &$(6.34c)^2+(-3.22c+5.9s)^2$   &$(2.00c+0.13s)^2+(-0.13c+4.23s)^2$  & \\    
$D_1'^0\rightarrow\gamma D^{*0}$    &384 &$(27.26c)^2+(19.35c-9.83s)^2$ &$(17.78c+0.15s)^2+(12.29c-6.13s)^2$ & \\
$D_1^+\rightarrow\gamma D^+$        &494 &$(5.49s+4.75c)^2$             &$(4.17s-0.60c)^2$               & \\
$D_1^0\rightarrow\gamma D^0$        &493 &$(8.78s+31.42c)^2$            &$(5.56s+18.78c)^2$              & \\
$D_1'^+\rightarrow\gamma D^+$       &494 &$(-5.49c+4.75s)^2$            &$(4.17c-0.60s)^2$               & \\
$D_1'^0\rightarrow\gamma D^0$       &498 &$(-8.90c+31.41s)^2$           &$(-5.62c+18.78s)^2$             & \\
$D_2^+\rightarrow\gamma D^{*+}$     &413 &15.0                          &6.49                            & \\
$D_2^0\rightarrow\gamma D^{*0}$     &412 &517                           &206                             & \\
\hline
$D_s^*\rightarrow\gamma D_s$        &139 &0.20                         &0.00                               & \\
$D_{s0}\rightarrow\gamma D_s^*$     &196 &6.85                         &0.16                               & \\
$D_{s1}\rightarrow\gamma D_s^*$     &322 &$(1.84s)^2+(0.99s+2.39c)^2$  &$(0.18s-0.07c)^2+(-0.44s+2.13c)^2$ & \\
$D_{s1}'\rightarrow\gamma D_s^*$    &388 &$(2.13c)^2+(-0.87c+3.62s)^2$ &$(0.24c-0.10s)^2+(0.64c+3.19s)^2$  & \\
$D_{s1}\rightarrow\gamma D_s$       &441 &$(2.68s+1.37c)^2$            &$(2.55s-1.21c)^2$                  & \\
$D_{s1}'\rightarrow\gamma D_s$      &503 &$(3.54c-1.12s)^2$            &$(3.33c+1.52s)^2$                  & \\
$D_{s2}\rightarrow\gamma D_s^*$     &420 &1.98                         &3.94                               & \\
\hline
\end{tabular}
\end{center}
\end{table}

%\begin{table}[h]
%\caption{Meson Radiative Decay rates (keV). $q_{\gamma}\rightarrow 0$ for the integral, global parameters have been used.}
%\label{RadTransTab}
%\begin{center}
%\begin{tabular}{|c|c|c|}
%\hline
%\hline
%mode & $q_\gamma$ (MeV) & nonrel \\
%\hline
%$D_0^+\rightarrow\gamma D^{*+}$     &364 &34                             \\
%$D_0^0\rightarrow\gamma D^{*0}$     &367 &837                            \\
%$D_1^+\rightarrow\gamma D^{*+}$     &381 &$33.5s^2$                      \\ 
%\hline
%\end{tabular}
%\end{center}
%\end{table}

Further complicating the analysis is the large difference seen between the nonrelativistic and
relativistic models (see, eg, $D^{+*}\to \gamma D^+$). This unfortunate circumstance is due to 
differing signs between the heavy and light quark impulse approximation subamplitudes. Employing
the full quark spinors leaves the heavy quark subamplitude largely unchanged, whereas the light
quark subamplitude becomes larger, thereby reducing the full amplitude. The effect appears to be 
at odds with the 
only available experimental datum ($D^* \to D \gamma$).  Clearly it would be useful to measure as
many radiative transitions as possible in these sectors to better evaluate the efficacy of these
(and other) models.

Once the decay model reliability has been established, ratios such as $\Gamma(D_{1} \to \gamma D^*)/\Gamma(D_1' \to \gamma D^*)$ and $\Gamma(D_1 \to \gamma D)/\Gamma(D_1\to \gamma D)$ will help determine
the $D_1$ mixing  angle.

%\begin{figure}
%\includegraphics[width= 6 true cm, angle=270]{Ds1decayRatio.ps}
%\caption{.....}
%\label{Ds1RatioFig}
%\end{figure}

\section{Discussion and Conclusions} % -------------------------------

A popular model of the $D_s$ mesons is based on an effective lagrangian description of
mesonic fields in the chiral and heavy quark limits\cite{chiral}. Deviations from these limits induce
mass splittings which imply that the axial--vector and  scalar-pseudoscalar mass differences are the
same. Since the premise of this idea has been questioned in Refs. \cite{HHreview, Bicudo}, it is
of interest to consider this mass difference in the present model. Splittings for the three
parameter sets considered above are shown in Table \ref{ChiralSplittingsTab}. Evidently, the 
chiral multiplet relationship holds to a very good approximation in both the $D$ and $D_s$ sectors and
is robust against variations in the model parameters. 

Nevertheless, the near equivalence of these mass differences must be regarded as an accident. Indeed,
the $B$ masses given in Table \ref{BTab}  indicate that this relationship no longer holds. It
would  thus be of interest to find P-wave open bottom mesons (especially scalars). These data
will distinguish chiral multiplet models from the model presented here and from more traditional
constituent quark models. For example, Godfrey and Isgur claim that the $B_0$ meson lies between
5760 and 5800 MeV; the $B_{s0}$ mass is 5840-5880 MeV, and the $B_{c0}$ mass is 6730-6770 MeV.
Of these, our $B_{s0}$ mass is predicted to be 65-105 MeV lower than the Godfrey-Isgur mass.

\begin{table}[h]
\caption{Chiral Multiplet Splittings (MeV).}
\label{ChiralSplittingsTab}
\begin{tabular}{l|cc}
\hline
\hline
params & $M(1^+(1/2^+)) - M(1^-)$ & $M(0^+) - M(0^-)$ \\
\hline
$D$ low & 411  &  412 \\
$D$ avg & 391  & 389 \\
$D$ high & 366 & 368 \\
\hline
$D_s$ low & 384 & 380 \\
$D_s$ avg & 373 & 370 \\
$D_s$ high & 349 & 346 \\
\hline
\end{tabular}
\end{table}

\begin{table}[h]
\caption{Low Lying Bottom Meson Masses (MeV)}
\label{BTab}
\begin{tabular}{l|cccccc}
\hline
\hline
flavour & $0^-$ & $1^-$ & $0^+$ & $1^+$ & $1^+$ & $2^+$ \\
\hline
$B$ & 5279 & 5322 & 5730 & 5752 & 5753 & 5759 \\
expt & 5279 & 5325 &  --  & $5724\pm 4 \pm 7$ &  --   &  $5748 \pm 12$ \\
\hline
$B_s$ & 5370 & 5416 & 5776 & 5803 & 5843 & 5852 \\
expt & 5369.6 & 5416.6 &  --  &  --   &  --   &  --  \\
\hline
$B_c$ & 6286 & 6333 & 6711 & 6746 & 6781 & 6797 \\
expt & 6286 & -- &  --  &  --   &  --   &  --  \\
\hline
\end{tabular}
\end{table}
The bottom flavoured meson spectra of Table \ref{BTab} have been
obtained with the `average' extended model parameters and $m_b = 4.98$ GeV. As with the open charm spectra, a flavour-dependent constant was fit to each pseudoscalar.
The second row reports recently 
measured P-wave $B$ meson masses\cite{D0}; these are in reasonable agreement with the predictions of the first row.

When these results are (perhaps incorrectly) extrapolated to light
quark masses, light scalar mesons are possible. Thus a simple $q\bar q$ interpretation of the 
enigmatic $a_0$ and $f_0$ mesons becomes feasible.

Finally, the work presented here may explain the difficulty in accurately computing the mass of the
$D_{s0}$ in lattice simulations. If the extended quark model is correct, it implies that
important mass and spin-dependent interactions are present in the one-loop level one-gluon-exchange
quark interaction. It is possible that current lattice computations are not sufficiently
sensitive to the ultraviolet behaviour of QCD to capture this physics. The problem is exacerbated
by the nearby, and presumably strongly coupled, $DK$ continuum; which requires simulations 
sensitive to the infrared behaviour of QCD. Thus heavy-light mesons probe a range of QCD scales
and make an ideal laboratory for improving our understanding of the strong interaction.

\acknowledgments

We are grateful to Wayne Repko for bringing several important references to our attention.
This work is supported by the U.S. Department of Energy under contract DE-FG02-00ER41135.

\end{document}